\documentclass[RNAAS]{aastex62}

\graphicspath{{./}{figures/}}

\shorttitle{Planet-Planet Tides in TRAPPIST-1}
\shortauthors{Wright}

\begin{document}

\title{Planet-Planet Tides in the TRAPPIST-1 System}

\correspondingauthor{Jason T.\ Wright}
\email{astrowright@gmail.com}
\author[0000-0001-6160-5888]{Jason T.\ Wright}
\affil{Department of Astronomy \& Astrophysics \\ and \\ Center for Exoplanets and Habitable Worlds \\ 525 Davey Laboratory \\
The Pennsylvania State University \\
University Park, PA, 16802, USA}

\keywords{planets and satellites: dynamical evolution and stability, planets and satellites: terrestrial planets}

\section{}
The star TRAPPIST-1 hosts a system of seven transiting, terrestrial exoplanets apparently in a resonant chain \citep{Gillon2016,Gillon17}, at least some of which are in or near the Habitable Zone.

Many have examined the roles of tides in this system, as tidal dissipation of the orbital energy of the planets may be relevant to both the rotational and orbital dynamics of the planets, as well as their habitability. \citep[e.g.][]{Vinson17,Auclair18,Valencia2018,Makarov2018,Papaloizou2018}
Generally, tides are calculated as being due to the tides raised on the planets by the star, and tides raised on the star by the planets.

I write this research note to point out a tidal effect that may be at least as important as the others in the TRAPPIST-1 system and which is so far unremarked upon in the literature: planet-planet tides.

At first blush, it would seem that planet-planet tides must not be relevant: the planets are many orders of magnitude less massive than the star, and so raise correspondingly smaller tides on each other than the star raises on them.

However, there are two considerations that bring the effects to near parity. The first is that the the planet-planet tides go from effectively zero (when two planets are far from each other) to their full strength during planetary conjunction, whereas if the planets are synchronously rotating, then the tides due to the star are relatively constant, modulated only by their very small orbital eccentricity.

The second is that the planets at conjunction are much closer to each other than they are to the star, and since the tidal bulge raised goes as the third power of the separation, this can be a large effect.

Consider the magnitude of the change in the tidal forces in the planets due to eccentricity tides and due to planet-planet interactions. The tidal strain $\varepsilon$ on a body $p$ scales with the mass $M$ and distance $d$ to the perturber as

\begin{equation}
\varepsilon_p = c_p f \frac{M}{d^3}
\end{equation}

\noindent where $c_p$ encapsulates all of the properties of the body $p$, and $f$ captures the fractional range over which the tidal forces vary. For tides raised by the star on a synchronously rotating planet with low eccentricity, $f=e$, and for planet-planet tides, the tidal forces vary from near zero to their full value, so we can take $f\sim 1$. We can then compare the strain induced by the star on planet $p$ to that induced by another planet $q$ with the ratio

\begin{equation}
\frac{\varepsilon_{p,q}}{\varepsilon_{p,*}} = \frac{m_q}{M_*|1-a_q/a_p|^3e_p}
\end{equation}

Using the values for $a$, $m$, $e$, and $M_*$ from \citet{Grimm2018} and \citet{VanGrootel2018},
I find that for every planet $p$ in the TRAPPIST-1 system there exists some other planet $q$ for which $\varepsilon_{p,q}/\varepsilon_{p,*} > 0.1$ during conjunction.  This suggests that in no case can planet-planet tides be neglected out-of-hand compared to stellar tides. Indeed, in the case of planet $g$ we have $\varepsilon_{g,f} / \varepsilon_{g,*} = 2.7$, meaning that planet-planet tides may dominate.

The effects of these tides are complex functions of the structure and spin states of the planets, which are not known. It is also nontrivial to compare the magnitudes of dynamical tides to eccentricity tides in this case: most existing treatments of dynamical tides are not immediately applicable to the TRAPPIST-1 system because they assume either mutual Keplerian motion between the bodies, or that they are gaseous stars or planets, or both. As a very rough first pass at plausibility, however, we can examine the rate of energy dissipation in the cases above, which scales as

\begin{equation}
\dot{E} = k \varepsilon^2\omega
\end{equation}

\noindent where $k$ is a function of not only the planet's properties, but also the nature of the interaction, and $\omega$ is the frequency of the perturbations. That is, $\omega$ is equal to the mean orbital motion $n_p = 2\pi/P_p$ of planet $p$ in the case of star-planet eccentricity tides (where $P$ is the orbital period), and $\omega$ is equal to the synodic frequency $|n_p - n_q|$ in the case of planet-planet tides.  

For all planets $p$ except $b$, there is another planet $q$ where $\varepsilon_{p,q}^2\omega_{p,q} / (\varepsilon_{p,*}^2\omega_{p,*}) > 1\%$, and in fact $\varepsilon_{g,f}^2\omega_{g,f} / (\varepsilon_{g,*}^2\omega_{g,*}) = 2.4$.

It is thus not obvious that planet-planet tides can be neglected in the TRAPPIST-1 exoplanetary system, especially the tides on planet $g$ due to planet $f$, if the planets are in synchronous rotation.

[{\bf Erratum:} After publication, I became aware of the work of \citet{LingamTides18}, who noted the potential importance of planet-planet tides on habitability by showing that  planet-planet tides between neighboring planets in the TRAPPIST-1 system have similar magnitudes to lunar tides on the Earth. The topic has thus not gone completely ``unremarked upon" as I stated in my research note.]

\acknowledgements

I thank Eric Ford and Andrew Shannon for helpful discussions on this topic.

\bibliography{references}

\end{document}